\documentclass[aps,prl,reprint,tightenlines,superscriptaddress,amsmath,amssymb]{revtex4-1}   

\usepackage{graphicx}  
\usepackage{bm}        
\usepackage{mathptmx}
\usepackage[backref=true,colorlinks=true,hyperfigures=true,citecolor=blue,linkcolor=red]{hyperref}
\usepackage{siunitx}

\begin{document}


\title{Nonlinear force balance at moving contact lines}


\author{Matthieu Roch\'{e}}
\author{Laurent Limat}
\author{Julien Dervaux}
\email{julien.dervaux@univ-paris-diderot.fr}
\affiliation{Laboratoire Mati\`{e}re et Syst\`{e}mes Complexes, Universit\'{e} Paris Diderot, CNRS UMR 7057, Sorbonne Paris Cit\'{e},
10 Rue A. Domon et L. Duquet, F-75013 Paris, France}

\begin{abstract}
The spreading of a liquid over a solid material is a key process in a wide range of applications. While this phenomenon is well understood when the solid is undeformable, its "soft" counterpart is still ill-understood and no consensus has been reached with regards to the physical mechanisms ruling the spreading of liquid drops over soft deformable materials. In this work we show that the motion of a triple line on a soft elastomer is opposed both by nonlinear localized capillary and visco-elastic forces. We give an explicit analytic formula relating the dynamic contact angle of a moving drop with its velocity for arbitrary rheology. We then specialize this formula to the experimentally relevant case of elastomers with Chasset-Thirion (power-law) type of rheologies. The theoretical prediction are in very good agreement with experimental data, without any adjustable parameters. Finally, we show that the nonlinear force balance presented in this work can also be used to recover the classical de Gennes model of wetting.
\end{abstract}

\maketitle

%


The spreading of a liquid on a solid surface is a physical process, called wetting, occurring in a myriad of practical and natural situations. It is therefore not surprising that the physics of wetting has become a cornerstone of technologies as diverse as nanoprinting \cite{Ru2014}, microfluidic \cite{Squires2005}, microfabrication \cite{Srinivasarao79},  water drainage \cite{Shahidzadeh2003}, dew and oil recovery \cite{BERTRAND2002}, paint, adhesive and coating industries \cite{Chaudhury1992,Sokuler2010,TakSing2011}, among others. It is also involved in a broad range of natural phenomena such as the self-cleaning property of lotus leaves \cite{Barthlott1997} or the spreading of mucus over epithelial surfaces \cite{Hills1985}. Interesting analogies have also been drawn between wetting and some morphology transitions in cellular aggregates \cite{Perezgonzalez2019}.

 In the century following the pioneering work of Young and Dupr\'{e}, the physics of wetting has been mainly concerned with the problem of liquid drops sitting or moving on "hard solids" \cite{DeGennes1985,Bonn2009}. Such materials are deformed only on atomic scales by capillary forces at the contact line where the liquid, the solid and the atmosphere meet. As a result, these deformations have been duly set-aside in the classical theory of wetting and the shape of small sessile droplets is ruled solely by a competition between the projections of the interfacial forces onto the solid surface, a result known as the Young-Dupr\'{e} equation.

\begin{figure}[t!]
\begin{center}
\includegraphics[scale=1]{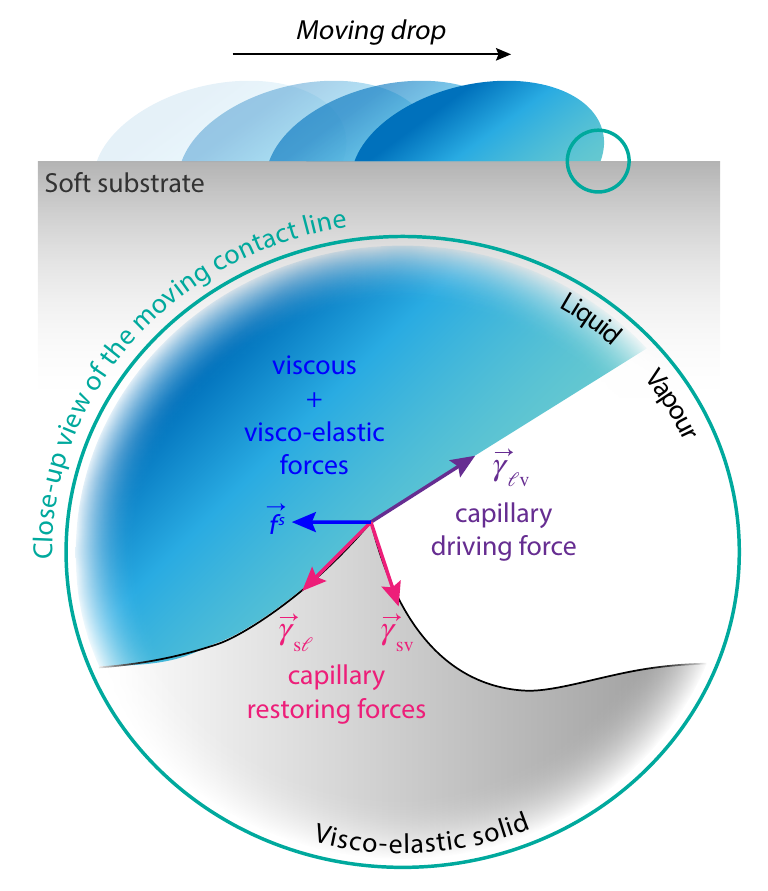}
\caption{Schematic representation of the problem. A liquid drop moves at constant velocity over a soft deformable substrate. At the contact line, the solid is deformed by the vertical component of the liquid surface tension $\vec{\gamma}_{\ell v}$ and develops a characteristic ridge. The motion of the contact line is opposed both by the visco-elastic force $\vec{f^{s}}$  and by restoring capillary forces $\vec{\gamma}_{sv}$ and $\vec{\gamma}_{s\ell}$ associated with the solid-vapour and solid-liquid interfaces.}
\label{fig1}
\end{center}
\end{figure}

Bikerman was apparently the first to realize, as early as 1957, that soft materials, which are significantly deformed by capillary forces at contact lines, were challenging the classical theory of wetting \cite{Bikerman57}.  However, further progresses had to wait for technological advances in microscopy and physico-chemistry.  In particular, the development of soft materials, such as gels and elastomers with controlled, highly tunable mechanical and surface properties, have allowed the detailed observation and characterization of a localized (typically micrometric) deformation in the solid, called a ridge, at the contact line between sessile drops and soft materials \cite{Pericet09,Style13,Park14}, illustrated in Fig. \ref{fig1}. Following these observations, many studies have investigated how this ridge, which may even persist following the removal of the drop \cite{Zhao2017}, modifies the shape of both the substrate and the sessile drop, with the aim of generalizing the classical theory of wetting to arbitrary materials \cite{Jerison2011,Limat2012,Style2012a,Bostwick2014,Lubbers2014,Dervaux2015}. On the dynamics side, it was shown that liquid drops move much more slowly over soft elastomeric layers than over rigid substrates \cite{Carre96}. This velocity reduction was shown to result from the dissipation of energy inside the elastomer, as a consequence of its geometric and visco-elastic properties \cite{Long96,Karpitschka2015,Zhao2018}. Surprising new effects such as stick-slip motion \cite{Kajiya2014}, spontaneous drop motion over substrate with a thickness gradient \cite{Style13b,Zhao2018} or attraction/repulsion between moving drops on soft substrates \cite{Karpitschka2016} have also been reported. 

One of the key question in the field of elastowetting is thus to understand the shape of this ridge and ultimately to predict the shape and contact angle of a liquid drop sitting or moving over a soft substrate. A corollary of this fundamental question is: \textit{if the Young-Dupr\'{e} equation does not hold anymore, what then is the relevant force balance equation at a contact line on a soft material ?} While the linear theory of elastowetting has suggested that the Neumann relation (the balance of surface tensions classically ruling the equilibrium of liquid drop over liquid layers) was also able to describe the wetting of liquid drop over soft elastic substrates, recent experiments have shown that this hypothesis does not hold for moving drops \cite{Zhao2018b} and leads to contradictory conclusions in static elastowetting \cite{Xu2017,Schulman2018,Masurel2019}. 

In order to answer this question, we have thus recently gone beyond the linear theory of elastowetting \cite{DePascalis2018,Masurel2019} and shown that the following nonlinear force balance (here generalized to arbitrary surface tensions and including possible external forces):

\begin{equation}
\vec{\gamma}_{\ell v} + \vec{\gamma}_{sv} + \vec{\gamma}_{s\ell} +  \vec{f^{ext}} = \vec{f^{s}} 
\label{eq:YDgeneralized}
\end{equation}

\noindent was able to explain in details the shape of \textit{static} elastocapillary ridges, for experimentally relevant value of the physical parameters. Here, $\vec{\gamma}_{\ell v}$, $\vec{\gamma}_{sv}$ and $\vec{\gamma}_{s\ell}$ are respectively the liquid-vapour, solid-vapour and solid-liquid capillary forces (per unit of length of the contact line). $ \vec{f^{ext}} $ represents additional external forces, other than surface tensions and visco-elastic stresses, that might act on the contact line. The force $ \vec{f^{s}} $ appearing on the right hand side of (\ref{eq:YDgeneralized})  is the force (per unit of length of the contact line) exerted by the deformed solid on the triple line. It can be written as:

\begin{equation}
 \vec{f^{s}}  = \lim_{\epsilon \rightarrow 0} \int_{\Gamma_{\epsilon}}\boldsymbol{\sigma} \cdot \vec{\nu} \, \mbox{d}\ell
\label{eq:eshelby0}
\end{equation}

\noindent where $\Gamma_{\epsilon}$ is a contour of radius $\epsilon$ enclosing the contact line in the deformed configuration, $\boldsymbol{\sigma}$ is the true (Cauchy) stress (to be defined below) and $\vec{\nu}$ is the outward unit vector normal to the contour. This relation is nonlinear because both $\boldsymbol{\sigma}$ and $\vec{\nu}$ depends on the deformation of the substrate. The force $\vec{f^{s}} $ typically does not vanish because the  elastocapillary ridge is a singular structure, or \textit{defect} in the language of Eshelbian mechanics. More precisely, because of the slope (and hence strain) discontinuity at the triple line, the ridge is akin to the terminal line of a generalized disclination \cite{DeGennes1995} whose strength can take any value between -1/2 and 1/2 as there is no underlying lattice structure, in contrast with disclinations in crystals. In a manner similar to disclinations, a localized force is thus exerted on the elastocapillary ridge whenever it is subjected to a deformation field (either its own deformation field or an additional external field, in which case this force is akin to the Peach-Koehler force \cite{Peach1950,Masurel2019}) or, as we shall see in the present paper, when the ridge moves. Equation (\ref{eq:YDgeneralized}) encompasses both the Young-Dupr\'{e} relation (for infinitely rigid substrate and no external forces $\vec{f^{ext}} =0$) as well as the Neumann relation ruling the equilibrium of liquid drop over liquid bath (in the limit of a substrate with vanishing shear and no external forces $\vec{f^{ext}} = \vec{f^{s}}=0$). In the general case however, the equilibrium angle of static droplets on soft substrates does not obey either of these two laws but satisfies equation (\ref{eq:YDgeneralized}). 

In this paper, we show that the generalized equation (\ref{eq:YDgeneralized}) can in fact also be exploited to predict the selection of the \textit{dynamic} contact angle of inviscid drops moving at constant velocities at the free surface of visco-elastic substrates. We show that both nonlinear localized capillary forces and visco-elastic stresses oppose the motion of the contact line. We give an explicit analytic formula relating the dynamic contact angle of a moving contact line with its velocity for arbitrary rheology. We then specialize this formula to the experimentally relevant case of elastomers with Chasset-Thirion (power-law) type of rheologies. The resulting theoretical prediction is in excellent agreement with experimental data, without any adjustable parameters. Finally, we show that equation (\ref{eq:YDgeneralized}) can also be used to recover the classical de Gennes model of dynamical wetting (for a viscous drop moving over a hard substrate) by choosing a contour $\Gamma_{\epsilon}$ located inside a liquid wedge moving at constant velocity.  

 \section{The case of an inviscid drop moving over a soft visco-elastic substrate}
 
\begin{figure}[t!]
\begin{center}
\includegraphics[width=\columnwidth]{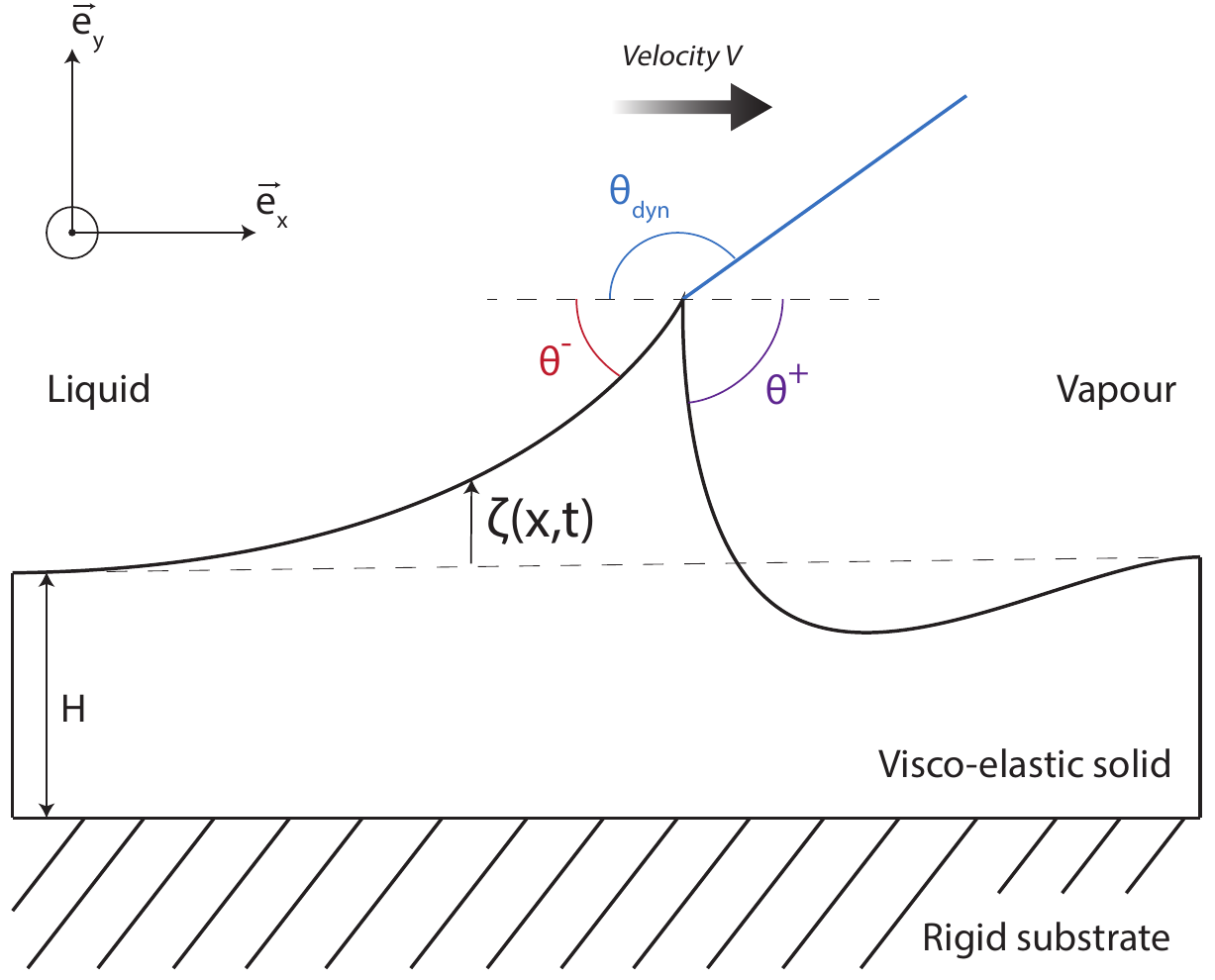}
\caption{Notation of the problem. A liquid drop with liquid surface tension $\gamma_{\ell}$ moves at constant velocity $V$ over a visco-elastic layer with initial thickness $H$ and surface tension $\gamma_s$. The drop is located on the left of the contact line and advances with a dynamic contact $\theta_{dyn}$ with the horizontal. The surface deformation of the visco-elastic layer is denoted by the function $\zeta(x,t)$. The slope of the interface has a jump discontinuity at the contact line. The angles $\theta^- = \vert \partial\zeta/\partial x\vert_{x=Vt^-}  \vert$ and  $\theta^+ = \vert \partial \zeta/\partial x \vert_{x=Vt^+} \vert$ are the (positive) slopes of the solid surface on each side of the triple line located at $x=Vt$.}
\label{fig2}
\end{center}
\end{figure}

Let us consider a single contact line moving with a velocity $V$ and a dynamic contact angle $\theta_{dyn}$ at the free surface of an intitially flat, incompressible, linearly visco-elastic layer with thickness $H$, as illustrated in Fig.\ref{fig2}. The reference state of the layer is described in cartesian coordinates by the region $-\infty < x < \infty$ and $-H<y<0$. The moving contact line deforms the layer and induces a displacement $\vec{u} = \{u_x(\vec{x},t),u_y(\vec{x},t)\}$ of material points initially located at $\vec{x} = \{x,y\}$. The liquid has a surface tension $\gamma_{\ell} \equiv \vert \vec{\gamma}_{\ell v} \vert $ and we assume for simplicity that the solid has a constant uniform surface tension $\gamma_s \equiv \vert \vec{\gamma}_{s v} \vert = \vert \vec{\gamma}_{s \ell} \vert$. We further assume that the stress $\boldsymbol{\sigma}(\vec{x},t)$ and strain $\boldsymbol{\epsilon}(\vec{x},t) = \partial \vec{u} / \partial \vec{x}$ tensors in the soft layer are related by the following general constitutive equation, valid for arbitrary linear visco-elastic materials: 

\begin{equation}
\boldsymbol{\sigma}(\vec{x},t) = \int_{-\infty}^t \mu(t-t') \frac{\partial \boldsymbol{\epsilon}}{\partial t'} \mbox{d}t' - p(\vec{x},t) \boldsymbol{I}
\label{eq:constit}
\end{equation}

\noindent where $\boldsymbol{I}$ is the identity matrix. The pressure $p(\vec{x},t)$ is the Lagrange multiplier associated with the incompressibility constraint. For stationary contact lines, the system is described by the following equilibrium equations:

\begin{equation}
\nabla \cdot \vec{u} =0 \,\,\,\, \mbox{and} \,\,\,\,  \nabla \cdot \boldsymbol{\sigma}= \vec{0}
\label{eq:equilibrium}
\end{equation}

We now write the boundary conditions needed to close the system. Motivated by several experimental setups, we assume that the lower face of the visco-elastic layer is bonded to an infinitely rigid substrate and thus at $y=-H$:

\begin{equation}
u_x(x,y=-H,t) = u_y(x,y=-H,t) =0 
\label{eq:BCbottom}
\end{equation}

Everywhere at the free surface ($y=0$), except at the contact line located at $x=Vt$, the normal projection of the visco-elastic stresses balances the Laplace pressure induces by the curved interface of the solid interface and we have:

\begin{equation}
\boldsymbol{\sigma} \cdot \vec{n} =\gamma_s \vec{n}\cdot (\nabla  \vec{n})  \,\,\,\, \mbox{at}  \,\,\,\, y=0 \,\,\,\, \mbox{and}  \,\,\,\, x \neq Vt
\end{equation}

At the contact line (at $y=0$ and $x=Vt$), we apply the  general force balance (\ref{eq:YDgeneralized}) specialized to the case of uniform solid surface tension and no external forces (i.e $\vec{f^{ext}} = \vec{0}$). This conditions reads, respectively along $\vec{e}_x$ and $\vec{e}_y$:

\begin{subequations}
\label{eq:BCtriple}
\begin{eqnarray}
-\gamma_{\ell} \cos{\theta_{dyn}} &=& \gamma_s \left\{ \cos{\theta^-} -\cos{\theta^+} \right\} + \vec{e}_x \cdot \vec{f^{s}} \label{eq:BCtriple1} \\
\gamma_{\ell} \sin{\theta_{dyn}} &=& \gamma_s \left\{ \sin{\theta^-} +\sin{\theta^+} \right\} + \vec{e}_y \cdot \vec{f^{s}} \label{eq:BCtriple2}
\end{eqnarray}
\end{subequations}

\noindent where $\theta^- = \vert \partial u_y/\partial x (Vt^-,0) \vert$ and  $\theta^+ = \vert \partial u_y/\partial x (Vt^+,0) \vert$ are the (positive) slopes of the solid surface on each side of the triple line located at $x=Vt$, as illustrated in Fig.\ref{fig2}. The force $\vec{f^{s}}$ exerted by the solid on the triple line is given by equation (\ref{eq:eshelby0}).

We now turn to the resolution of the boundary-value problem (\ref{eq:constit}-\ref{eq:BCtriple}). Within the framework of a linear theory, the slopes of the deformed solid surface must be small compared to unity for consistency. This condition will be verified when $\gamma_{\ell}/2\gamma_s \ll 1$. Taking this quantity as a small parameter, we may seek solutions of the boundary-value problem (\ref{eq:constit}-\ref{eq:BCtriple}) in the form:

\begin{equation}
\{\vec{u},p\} = \sum_{i=1}^{\infty} \{\vec{u}^i,p^i\} \left(\frac{\gamma_{\ell}}{2 \gamma_s}\right)^i
\label{eq:solutiondev}
\end{equation}

In that case a double Fourier transform with respect to both time and space yields the following first-order solution for the surface deflection $\zeta(x,t) \equiv (\gamma_{\ell}/2\gamma_s) u^1_y(x,y=0,t)$:

\begin{equation}
\zeta(x,t)= \frac{1}{2\pi }\int_{-\infty}^{\infty}\mbox{d}k e^{ik(x-Vt)} \hat{\zeta}(k) 
\label{eq:solution0}
\end{equation}

\noindent with


\begin{equation}
\hat{\zeta}(k) = \frac{\gamma_{\ell} \sin{\theta_{dyn}}}{\gamma_s} \left[k^2 +F(k) \right]^{-1} 
\label{eq:solution}
\end{equation}

\noindent and

\begin{equation}
F(k)=  \left[\frac{2H^2k^2+\cosh(2Hk)+1}{\sinh(2Hk)-2Hk}\right]\frac{2k\hat{\mu}(-kV)}{\gamma_s}
\label{eq:solution2}
\end{equation}


\noindent where we have used the particular definition of $\hat{\mu}(\omega)$:

\begin{equation}
\hat{\mu}(\omega) = i \omega \int_0^{\infty} \mu(t) e^{-i\omega t}\mbox{d}t
\label{eq:CTfour}
\end{equation}

The solution (\ref{eq:solution0}-\ref{eq:solution2}) above satisfies the boundary-value problem (\ref{eq:constit}-\ref{eq:BCtriple}) at first order in $\gamma_{\ell}/2\gamma_s$ provided that  $\cos{\theta_{dyn}} =\mathcal{O}(\gamma_{\ell}/2\gamma_s)$ or, equivalently, that $\theta_{dyn} = \pi/2 + \mathcal{O}(\gamma_{\ell}/2\gamma_s)$. In particular, let us underline that both components of the force $\vec{f^{s}}$ vanish at first-order. The solution  (\ref{eq:solution0}-\ref{eq:solution2}) however, does not specifies the dynamic contact angle beyond the zeroth-order approximation ($\theta_{dyn} = \pi/2$), it only provides the deformed profile of the interface in response to a vertical localized force of magnitude $\gamma_{\ell}\sin{\theta_{dyn}}$. The departure of the dynamic contact angle $\theta_{dyn}$ from $\pi/2$ results from higher order contributions that we now consider.

Under the condition that $\cos{\theta_{dyn}} =\mathcal{O}(\gamma_{\ell}/2\gamma_s)$ and using the solution (\ref{eq:solution0}-\ref{eq:solution2}), it can be verified that all the contributions coming form this first-order solution in equation (\ref{eq:BCtriple2}) are in fact at most of order $(\gamma_{\ell}/2\gamma_s)^3$ (after dividing equation (\ref{eq:BCtriple2}) by $\gamma_s$). On the other hand, all the terms in equation (\ref{eq:BCtriple1}) coming from the first-order solution are of order $(\gamma_{\ell}/2\gamma_s)^2$ (again after dividing equation (\ref{eq:BCtriple1}) by $\gamma_s$). As a consequence, the solution (\ref{eq:solution0}-\ref{eq:solution2}) will be a solution of (\ref{eq:constit}-\ref{eq:BCtriple}) up to second-order (i.e the next non-zero correction in (\ref{eq:solutiondev}) will be the third order term  $\{\vec{u}^3,p^3\}$ only if equation (\ref{eq:BCtriple1}) holds at second order.

Using the solution (\ref{eq:solution0}-\ref{eq:solution2}), the restoring capillary forces (the first term in the right-hand side of (\ref{eq:BCtriple1})) can be rewritten in the small slope limit as:
 
 \begin{eqnarray}
\gamma_s \left\{ \cos{\theta^-} -\cos{\theta^+} \right\}  &=& \frac{\gamma_{s}}{2} \{(\theta^+)^2 - (\theta^-)^2\} \nonumber \\
&=& \frac{\gamma_{\ell}\sin{\theta_{dyn}}}{2} \left( \theta^+ - \theta^-\right) \nonumber \\
&=& \frac{\gamma_{\ell}\sin{\theta_{dyn}}}{2 \pi} \mathfrak{R} \left[\int_{-\infty}^{\infty} -i k \hat{\zeta}(k)\mbox{d}k \right]\,\,\,\,\,\,
\label{eq:capfourier}
\end{eqnarray}

\noindent where the symbol $\mathfrak{R}$ stands for the real part. Note that $\left( \theta^+ - \theta^-\right)/2$ physically represents the rotation of the elastocapillary ridge. The horizontal restoring visco-elastic force acting at the tip of the ridge (the second term in the right-hand side of (\ref{eq:BCtriple1})) is found by injecting the solution (\ref{eq:solution0}-\ref{eq:solution2}) into equation (\ref{eq:eshelby0}). Its first non-zero contribution can be transformed into the following integral:

\begin{equation}
\vec{e}_x \cdot \vec{f^{s}}= \int_{-\infty}^{\infty}\sigma_{xx}^1 \frac{\partial \zeta}{\partial x} \mbox{d}x  
\label{eq:eshelby1}
\end{equation}

\noindent where $\sigma_{xx}^1$ is the first-order stress field associated with the first-order solution $\{\vec{u}^1,p^1\}$. Using Parseval's theorem, we can rewrite equation (\ref{eq:eshelby1}) in the frame of the moving ridge as:

\begin{equation}
\vec{e}_x \cdot \vec{f^{s}} =\frac{1}{2 \pi} \mathfrak{R} \left[ \int_{-\infty}^{\infty} i k^2 \hat{\mu}(kV)\hat{\zeta}(k)\hat{\zeta}(-k) \mbox{d}k \right]
\label{eq:eshelby-fourier}
\end{equation}

Note that expressions  (\ref{eq:capfourier}) as well as (\ref{eq:eshelby-fourier}), after a division by $\gamma_s$, are both of the same order $\mathcal{O}(\gamma_{\ell}/2\gamma_s)^2$. This shows that the Neumann construction does not hold in dynamic elastowetting and that the capillary driving force (i.e the left-hand side of (\ref{eq:BCtriple1})) is resisted not only by the restoring capillary force (\ref{eq:capfourier}) but also by the visco-elastic force (\ref{eq:eshelby-fourier}). Rearranging the terms in (\ref{eq:BCtriple1})), we obtain the following nonlinear force balance at the contact line:
 
 \begin{equation}
 \frac{\cos{\theta_{dyn}}}{\sin^2{\theta_{dyn}}} = \frac{\gamma_{\ell}}{2\pi \gamma_s}  \mathfrak{R} \left[ \int_{-\infty}^{\infty}\frac{i (k^2+F(-k)-k\hat{\mu}(kV)/\gamma_s)}{(k^2+F(-k))(k^2+F(k))}\right]
 \label{eq:nonlinearbalance}
\end{equation}

Equation (\ref{eq:nonlinearbalance}) gives the relation between the dynamic contact angle and the velocity of a contact line moving on a substrate of arbitrary rheology and thickness. We now compare Eq. (\ref{eq:nonlinearbalance}) to experimental data and more specifically to the motion of contact line on elastomeric substrates behaving according to the Chasset-Thirion model for which the complex frequency-dependent visco-elastic modulus $\mu(\omega)$ is:

\begin{equation}
\mu(\omega) = \mu_0 (1+(i\omega\tau)^m)
\label{eq:CT}
\end{equation}

\begin{figure*}[t!]
\begin{center}
\includegraphics[width=\textwidth]{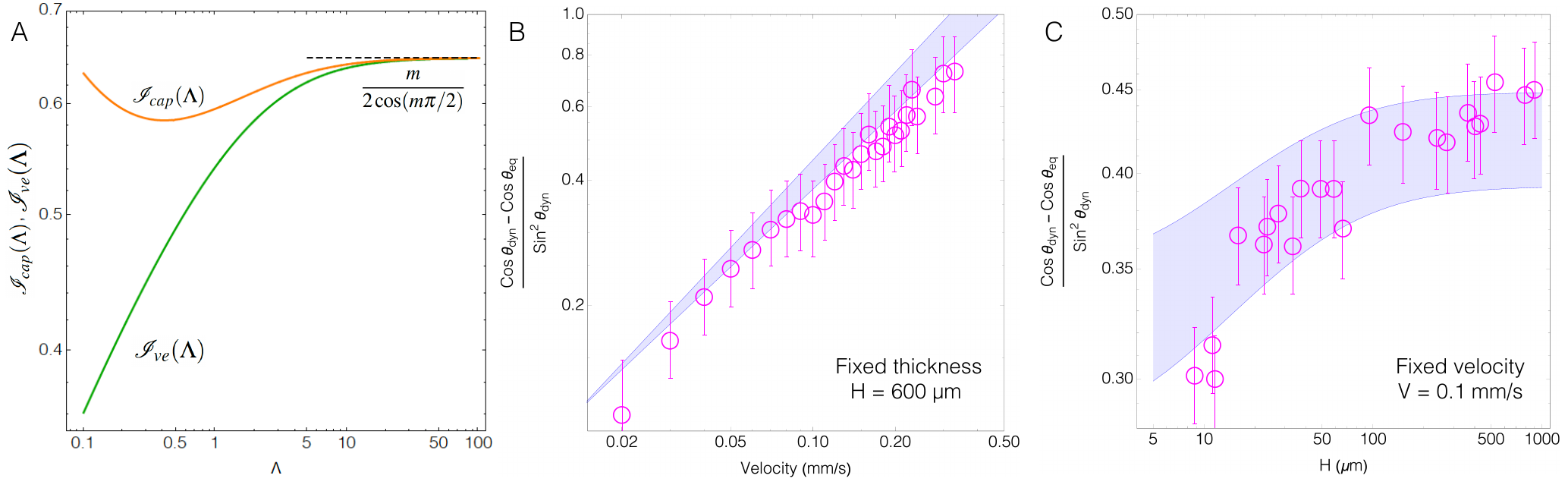}
\caption{Results of the theoretical model and comparison with experimental data. A: Plot of the dimensionless functions $\mathcal{I}_{cap}(\Lambda)$ and $\mathcal{I}_{ve}(\Lambda)$, defined respectively by Eq.(\ref{eq:nodimcap}) and Eq.(\ref{eq:nodimve}), that capture the dependance of the dynamic contact angle $\theta_{dyn}$ on the aspect ratio $\Lambda = H/\ell_s$ of the problem. Both functions converge to $m/(2\cos(m\pi/2))$ in the limit $\Lambda \rightarrow \infty$. B and C: Comparison of experimental data (magenta circles) with Eq.(\ref{eq:nonlinearbalanceCT2}) (blue shaded area) at a fixed thickness of $600\mu$m (B) and fixed velocity $0.1$mm/s (C). The data are taken from \cite{Zhao2018} and the values of the physical parameters are $\gamma_{\ell} = 72\pm2$ mN/m, $\gamma_{s} = 42\pm2$ mN/m, $\mu_0=1085\pm124$, $\tau=15.4\pm0.4$ ms and m$=0.66\pm0.04$. The width of the theoretical predictions in B and C reflects the uncertainties in the values of the physical parameters.}
\label{fig:experiment}
\end{center}
\end{figure*}

In the case where the velocity of the contact line is small enough (i.e for $V \tau/\ell_s \ll1$), the dynamic force balance takes the simple form:

 \begin{equation}
- \frac{\cos{\theta_{dyn}}}{\sin^2{\theta_{dyn}}} = \frac{\gamma_{\ell}}{\gamma_s}\left( \frac{V \tau}{\ell_s}\right)^m \mathcal{I}\left(\frac{H}{\ell_s}\right) \label{eq:nonlinearbalanceCT}
\end{equation}

\noindent where $\ell_s = \gamma_s/(2\mu_0)$ is the elastocapillary length while the dimensionless function $\mathcal{I}$ captures the dependence of the scaling on the reduced thickness $\Lambda=H/\ell_s$ of the problem. It can be written as the sum of the capillary and the visco-elastic contributions:

 \begin{equation}
\mathcal{I}(\Lambda)  =\mathcal{I}_{cap}(\Lambda) + \mathcal{I}_{ve}(\Lambda) 
\end{equation}

\noindent where

 \begin{equation}
\mathcal{I}_{cap}(\Lambda) =  \int_0^{\infty} \frac{k^m G(k\Lambda) \sin{(m\pi/2)}}{\pi(k+G(k\Lambda))^2}\mbox{d}k
\label{eq:nodimcap}
\end{equation}

\noindent and

 \begin{equation}
\mathcal{I}_{ve}(\Lambda) =  \int_0^{\infty} \frac{k^m \sin{(m\pi/2)}}{\pi(k+G(k\Lambda))^2}\mbox{d}k
\label{eq:nodimve}
\end{equation}

\noindent with

\begin{equation}
G(z)=  \left[\frac{2z^2+\cosh(2z)+1}{\sinh(2z)-2z}\right]
\label{eq:solutionG}
\end{equation}

In the limit of an infinitely thick substrate ($\Lambda \rightarrow \infty$) these integrals converge to the same limit $m/(2\cos(m\pi/2))$, yielding the result:

 \begin{equation}
\mathcal{I}_ \infty \equiv \lim_{\Lambda \rightarrow \infty} \mathcal{I}(\Lambda)  = 2  \lim_{\Lambda \rightarrow \infty}\mathcal{I}_{cap}(\Lambda) = \frac{m}{\cos(m\pi/2)}
\end{equation}

On an infinitely thick substrate, the restoring capillary forces and the visco-elastic force thus contribute equally to the horizontal force balance at the moving ridge. On the other hand, those two contributions differ for finite values of the aspect ratio $\Lambda$, as shown in Fig.\ref{fig:experiment}-A. In particular, the restoring visco-elastic force follows the scaling $\sim \Lambda^{3(1-m)/4}$ at small thickness, in agreement with a previous analysis \cite{Zhao2018}. Before proceeding to a comparison of the model with experiments, it should be noted that available experimental data on the dynamics of contact lines on soft solids are not performed with liquids in neutral wetting situation (\textit{i.e} with $\theta_{eq} = \pi/2$). In order to compare the model with experimental data, we shall therefore make the assumption that Eq.(\ref{eq:nonlinearbalanceCT}) still holds in the vicinity of an arbitrary equilibrium contact angle.  Under this assumption, we have:

 \begin{equation}
\frac{\cos{\theta_{eq}}-\cos{\theta_{dyn}}}{\sin^2{\theta_{dyn}}} = \frac{\gamma_{\ell}}{\overline{\gamma_s}}\left( \frac{V \tau}{\overline{\ell_s}}\right)^m \mathcal{I}\left(\frac{H}{\overline{\ell_s}}\right) \label{eq:nonlinearbalanceCT2}
\end{equation}

\noindent where $\overline{\gamma_s}$ must be understood as the average value between $\gamma_{sv}$ and $\gamma_{s\ell}$ and $\overline{\ell_s}= \overline{\gamma_s}/(2\mu_0)$. Note that while the present analysis has been performed for an advancing contact line, it also holds for a receding contact line, simply by changing the sign of the left hand sign in Eq.(\ref{eq:nonlinearbalanceCT2}). This theoretical prediction is compared to experimental data taken from \cite{Zhao2018} in Fig.\ref{fig:experiment}-B and C, using the values of the physical parameters given in  \cite{Zhao2018}. Despite the many approximations of our model, and in particular the fact that the parameter $\gamma_{\ell}/(2 \overline{\gamma_s}) \sim 0.9$ is not much smaller than unity, the agreement between theory and experiments is excellent, without any adjustable parameters.


 \section{the case of a viscous drop moving over a rigid substrate}
 We now consider the case of a viscous drop moving on a rigid, purely elastic, substrate. In that case, equation (\ref{eq:BCtriple}) can again be used to find the force balance at the contact line. In the limit of an infinitely rigid substrate, the angles $\theta^-$ and $\theta^+$ vanish and the vertical capillary traction is balanced solely by the elastic stresses \cite{Dervaux2015}. In the horizontal direction we are thus left with the relation (\ref{eq:BCtriple1}):
 
 \begin{equation}
 -\gamma_{\ell v} \cos{\theta_{dyn}} + \vec{e}_x \cdot \vec{f^{ext}} = \gamma_{s v} -\gamma_{s \ell} +\vec{e}_x \cdot \vec{f^{s}} 
 \end{equation}
 
 \noindent but now the localized force $\vec{f^{s}}$  must be evaluated by using a contour inside the viscous fluid. As shown by Moffat \cite{moffatt_1964}, in the vicinity of a liquid wedge defined by a free surface moving at velocity $V$ with respect to a solid surface, the fluid flow is self-similar in polar coordinates $(r,\theta)$ and can be derived from the stream function $\psi = V r f(\theta)$. More precisely, we have for the velocity component $(v_r,v_{\theta})$ of the fluid flow:
 
  \begin{equation}
v_r= \frac{1}{r}\frac{\partial \psi}{\partial \theta} = V f'(\theta), \,\,\,\, v_{\theta}= -\frac{\partial \psi}{\partial r} = -V f(\theta)
 \end{equation}
 
\noindent where $f(\theta)$ is:

 \begin{equation}
f(\theta) = \frac{\theta \cos\theta \sin\theta_{dyn} - \theta_{dyn} \cos\theta_{dyn}\sin{\theta}}{\sin\theta_{dyn} \cos\theta_{dyn} -\theta_{dyn}}
 \end{equation}
 
 \noindent while the pressure field $p(r,\theta)$ is given by:
 
 \begin{equation}
p(r,\theta) = \frac{\eta}{r} \frac{4 V \cos\theta \sin\theta_{dyn}}{\sin(2\theta_{dyn})  - 2\theta_{dyn}}
 \end{equation}

\noindent where $\eta$ is the dynamic viscosity. Here the free surface is located at $\theta=0$ while the solid substrate is located at $\theta=-\theta_{dyn}$. Note that this solution requires the presence of a normal surface pressure $4V\mu\sin{\theta_{dyn}}/(r(\sin(2\theta_{dyn})-2\theta_{dyn}))$ at the free surface in order to keep the interface flat. Using the solution above for the velocity and pressure fields, one can easily calculate the horizontal projection of the viscous force as

 \begin{equation}
\vec{e}_x \cdot \vec{f^{s}}  = -4 V \eta \frac{\sin^2\theta_{dyn}}{\sin(2\theta_{dyn})  - 2\theta_{dyn}}
 \end{equation}
 
 A striking feature of this result is that, although the pressure and stress fields diverge at the corner with an essential singularity in $\sim r^{-1}$, the resulting viscous force at the corner is finite (and in fact independent of the choice of contour). In addition to the viscous force, there is also a contribution at the tip of the moving wedge coming from the (divergent) surface pressure $4V\mu\sin{\theta_{dyn}}/(r(\sin(2\theta_{dyn})-2\theta_{dyn}))$ at the free surface needed to keep the interface flat. The usual approximation in dynamical wetting consists in integrating this contribution between a microscopic scale $x_{min}$ and a macroscopic scale $x_{max}$ in order to obtain the force $\vec{f^{ext}}$, and thus:
 
  \begin{equation}
\vec{e}_x \cdot \vec{f^{ext}}  = 4 V \eta \frac{\sin^2\theta_{dyn}}{\sin(2\theta_{dyn})  - 2\theta_{dyn}} \log \left( \frac{x_{max}}{x_{min}} \right)
 \end{equation}
 
  Once summed with the viscous force at the tip, we obtain the following force balance at the tip of the moving wedge:

 \begin{equation}
\cos\theta_{dyn}-\cos\theta_{eq}  = 4 \mathcal{C}_a \frac{\sin^2\theta_{dyn}}{\sin(2\theta_{dyn})  - 2\theta_{dyn}}   \left[ 1 + \log \left( \frac{x_{max}}{x_{min}} \right) \right]
 \end{equation}
 
 \noindent where $\mathcal{C}_a = \eta V/ \gamma_{\ell}$ is the capillary number. Interestingly, it should be noted that in the small angle limit ($\theta_{eq}\ll$,$\theta_{dyn}\ll$), we recover the classical de Gennes model of wetting:
 
  \begin{equation}
\theta_{dyn}\left(\theta_{dyn}^2-\theta_{eq}^2 \right) = 6 \mathcal{C}_a \left[ 1 + \log \left( \frac{x_{max}}{x_{min}} \right) \right]
 \end{equation}
 
Note that this scaling is a consequence of the flat interface approximation used in this section, which requires an external force. In a more general context, the divergent viscous stress induces a bending of the interface and leads to the Cox-Voinov relation. 
 
\section{Conclusion}
 
We have shown in this paper that the generalized force balance (\ref{eq:YDgeneralized}) was able to describe accurately, and without any adjustable parameters, the dependance of the dynamic contact angle on both the velocity of a contact line as well as the material and geometric properties of the soft substrate. We have demonstrated that the motion of the triple line is opposed by both visco-elastic and capillary forces, the two contributions being equal in the limit of an infinitely thick substrate. This observation explains why, in a model based purely on capillarity, such as the Neuman force balance, it was found necessary to redefine the surface tension (roughly by dividing it by $\sim2$) in order to fit the experimental data \cite{Karpitschka2015}. Although this was attributed to a finite slope effect, the model presented here shows that this is simply a consequence of the substrate visco-elasticity. In addition, we have also shown that the same equation allows to recover the classical de Gennes model for the spreading of a viscous drop on a rigid substrate. Therefore, although the experimental data were limited to slowly moving drops, for which the dissipation mostly occurs in the visco-elastic substrate, we expect the framework presented here to be able to describe much more general situations, for which the dissipation occur simultaneously in both the solid and the liquid phases. In addition, the generalized force balance (\ref{eq:YDgeneralized}) can also be applied to the yet largely unexplored case of the spreading of a viscous drop over a viscous liquid bath.
 

\section{acknowledgments}
We thank ANR (Agence Nationale de la Recherche) and CGI (Commisariat \`{a} l'Investissement d'Avenir) are gratefully acknowledged for their financial support through the GELWET project (ANR-17- CE30-0016), the Labex SEAM (Science and Engineering for Advanced Materials and devices - ANR 11 LABX 086, ANR 11 IDEX 05 02) and through the funding of the POLYWET and MMEMI projects.

\bibliographystyle{apsrev}
\bibliography{dynamicforcebalance}

\end{document}